# Hardware Efficient WiMAX Deinterleaver Capable of Address Generation for Random Interleaving Depths

Omar Rafique[#1], Gangadharaiah S.L.[#2]

[#1] *PG Scholar, Dept. of Electronics & Communication Engineering, M.S.R.I.T, Bangalore,Karnataka, India*
[#2] *Asst. Professor, Dept. of Electronics & Communication Engineering, M.S.R.I.T, Bangalore,Karnataka, India*

***Abstract* -- The variation in the prescribed modulation schemes and code rates for WiMAX interleaver design, as defined by IEEE 802.16 standard, demands a plethora of hardware if all the modulation schemes and code rates have to be unified into a single electronic device. Add to this the complexities involved with the algorithms and permutations of the WiMAX standard, invariably dependent on floor function which is extremely hardware inefficient. This paper is an attempt towards removing the complexities and excess hardware involvement in the implementation of the permutations involved in Deinterleaver designs as defined by IEEE 802.16**

## I INTRODUCTION

The significance of this paper lies in the fact that the constant evolution of communication standards always poses a challenge to the development of a unifying algorithm standard. This paper is an attempt to incorporate random code rates as a possible convenience for future implementations of WiMAX deinterleaver design and at the same time focusing on area efficiency of the overall circuit by resource sharing. This paper is an amendment of the designs suggested by [1].

WiMAX (Worldwide Interoperability for Microwave Access) , defined by IEEE 802.16e standard, was created in 2001 and is capable of delivering up to 70 megabit data rates per second with the added advantage of mobility[4].

## II THE WiMAX MODEL

The WiMAX model mainly consists of the following blocks: (Figure 1)

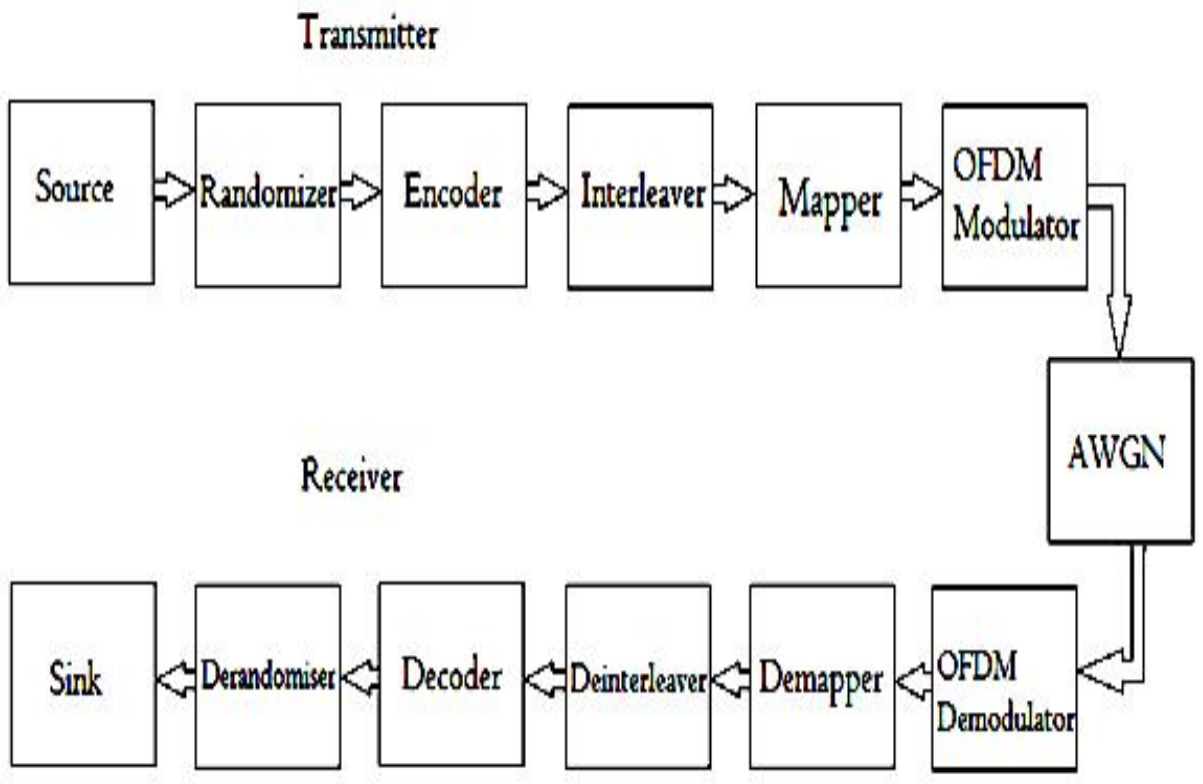

Fig. 1 The WiMAX Block Diagram

The randomizer eliminates a long sequence of zeros and ones so that synchronization is not lost. It works on bit by bit basis. Encoding is used for forward error correction where additional redundancy bits are added to the output of the randomizer. Interleaver is used for protection against burst errors which can make a sequence of consecutive bits erroneous, thus making it difficult for the error correcting codes to correct this long sequence of consecutive errors. WiMAX uses Reed Solomon Codes and its error correcting capacity is 8 bits. If there are more than 8 consecutive bits in error than RS code will not be able to correct them. It is the job of the interleaver to break this sequence of consecutive erroneous bits and make it possible to correct errors by RS codes. The mapper maps the incoming bits onto a constellation. The governing equation for WiMAX interleaver/deinterleaver is a two step permutation and is defined by IEEE 802.16 standard is given by [3],[4]:

$$m_k = \frac{N_{cpbs}}{d} \times (k\%d) + \left\lfloor \frac{k}{d} \right\rfloor \qquad (1)$$

$$j_k = s \times \left\lfloor \frac{m_k}{s} \right\rfloor + \left( \left( m_k + N_{cpbs} - \left\lfloor d \times \frac{m_k}{N_{cpbs}} \right\rfloor \right) \% s \right) \qquad (2)$$

The deinterleaver performs inverse operations and the permutations for it are defined as follows:

$$m_j = s \times \left\lfloor \frac{j}{s} \right\rfloor + \left( j + \left\lfloor \frac{d.j}{N_{cpbs}} \right\rfloor \right) \% s \qquad (3)$$

$$k_j = d.m_j - (N_{cpbs} - 1) \times \left\lfloor \frac{d.m_j}{N_{cpbs}} \right\rfloor \qquad (4)$$

Equations (3) and (4) define a two level permutation and $j$ is the index of the received bit within a block of $N_{cpbs}$ bits. The first equation is for mapping the adjacent codded bits onto non adjacent subcarriers of the OFDM modulation scheme and second equation and the second permutation maps them alternately onto less or more significant bits of the constellation thus avoiding long runs of less reliable bits. The letter *'d'* represents the number of rows of the block deinterleaver and may be chosen as 12 or 16. We have preferred d = 16 in this paper as 16 is a power of 2 and division with any number other than the power of 2 is not feasible for FPGA implementation. '*k*' varies from 0 to *N* and $M_k$ is the output of the first equation and $J_k$ is the output of the second equation. The parameter '*s*' is defined as max{1, $N_{cpbs}/2$} , where $N_{cpbs}$ stands for numner of codded bits per symbol of the OFDMA modulation scheme. The floor fuction has been represented as $\lfloor \ \rfloor$.

## III EXISTING METHODS

Not much work has been done regarding the hardware efficiency of WiMax deinterleaver, owing to it being a relatively young technology in the field of high speed , broadband communication. Two prominent sources are Upadhyaya and Sanyal [1] and Asghar and Liu [2]. [2] suffers from some mathematical complexities and is not clearly explained whereas [1], though being a simple approach can be further simplified to incorporate random code rates for future developments and at the same time incorporation further hardware efficiency. The following sections consist of the work we have done, the results and discussions and finally the conclusion.

## IV PROPOSED METHOD

An alternate algorithm to generate the deinterleaver addresses by eliminating the floor function has been obtained by [1]. The addresses generated for different code rates and modulations are shown in Tables I,II,III. These are the addresses generated by using the permutation formulae given by IEEE 802.16 standard for WiMAX deinterleaver. Here the value of 'd' has been chosen as 16 , though other values are also possible (e.g.12). This is done keeping in the the final synthesis problems i.e. Xilinxs does not allow division by a number other than the power of two.

The algorithm used to generate the above addresses, thus bypassing the floor function of the permutation formulae defined by IEEE are defined as in figure 2. Here $j$ (0 to d-1) gives us the row number and $i$ (0 to $(N_{cpbs}/d)$-1) gives the column number. $k_n$ represents the deinterleaver addresses. We have modified the implemented circuitry for the above algorithm keeping in view the optimality of the final design in terms of generalization of the address generator to accept random code rates and the final hardware efficiency. The designing process has been divided into four modules. The first three modules deal with the designs of QPSK, 16-QAM and 64-QAM address generators and the final module deals with the combination of these three modules using resource sharing. A divider and a subtractor (minus one) circuit has been used as a common input module for all the three types of modulations and resource sharing has been done in terms of the blocks common to all the three modulation schemes. The divider and a subtractor (minus one) combination at the input of the modulation blocks makes it possible to accept random code rates and also eliminates the use of multiple multiplexer modules as done by [1] and as it is common to all the three modules, we can use only one divider and subtractor module instead of three, thereby further reducing the hardware. The block diagram representation has been shown in figures III,IV,V and VI for QPSK, 16-QAM and 64-QAM and the final deinterleaver block.

TABLE I :
QPSK Addresses (First 5 rows)

| **Modulation/ $N_{cpbs}$** | **Deinterleaver addresses for first 5 rows** | | | | |
|---|---|---|---|---|---|
| **QPSK/96 bits** | 0 | 16 | 32 | 48 | 64 |
| | 1 | 17 | 33 | 49 | 65 |
| | 2 | 18 | 34 | 50 | 66 |
| | 3 | 19 | 35 | 51 | 67 |
| | 4 | 20 | 36 | 52 | 68 |

TABLE III :
16-QAM Addresses (First 5 rows)

| **Modulation/ $N_{cpbs}$** | **Deinterleaver addresses for first 5 rows** | | | | |
|---|---|---|---|---|---|
| **16-QAM/192 bits** | 0 | 16 | 32 | 48 | 64 |
| | 17 | 1 | 49 | 33 | 81 |
| | 2 | 18 | 34 | 50 | 66 |
| | 19 | 3 | 51 | 35 | 83 |
| | 4 | 20 | 36 | 52 | 68 |

TABLE IIII :
64-QAM Addresses (First 5 rows)

| **Modulation/$N_{cpbs}$** | **Deinterleaver addresses for first 5 rows** | | | | |
|---|---|---|---|---|---|
| **64-QAM/596 bits** | 0 | 16 | 32 | 48 | 64 |
| | 17 | 33 | 1 | 65 | 81 |
| | 34 | 2 | 18 | 82 | 50 |
| | 3 | 19 | 35 | 51 | 67 |
| | 20 | 36 | 4 | 68 | 84 |

## V RESULTS

Xilinx ISE Design Suite 12.3 was used to implement the modified version of the deinterleaver. The results are tabulated in Table IV where comparison has been drawn between the LUT based technique, the technique proposed by [1 ]and our proposed scheme. It was observed that significant reduction is obtained in hardware utility and at the same time the operating frequency also increased. The Simulation is shown in Figure VII, which shows the addresses generated for code rate zero and interleaving depth (Ncpbs) equal to 96.

## VI CONCLUSION

Apart from successfully optimizing the hardware and operating frequency requirements by complexities arising from the presence of the floor function in the permutation formulae defined by IEEE 802.16, our proposed method has also made the deinterleaver a generalized circuit in terms of accepting any range of Interleaving Depth (Ncpbs) thus rendering it as a flexible in terms of future developments.

## REFERENCES

[1] Bijoy Kumar Upadhyaya , and Salil Kumar Sanyal, “Efficient FPGA Implementation ofAddress Generatorfor WiMAX Deinterleaver,” IEEETRANSACTIONS ONCIRCUITS AND SYSTEMS—II: EXPRESS BRIEFS,VOL. 60, NO. 8, AUGUST 2013

[2] R. Asghar and D. Liu, “2D realization of WiMAX channel interleaver for efficient hardware implementation,” in *Proc. World Acad. Sci. Eng. Technol.*, Hong Kong, 2009, vol. 51, pp. 25–29.

*[3]* IEEE Standard for Local and Metropolitan Area Networks — Part 16: Air Interface for Fixed Broadband Wireless Access Systems—Amendment 2, IEEE Std. 802.16e - 2005, 2005.

*[4]* M. N. Khan and S. Ghauri, “The WiMAX 802.16e Physical layer model,” in *Proc. IET Int. Conf.* Wireless, Mobile Multimedia Netw., Mumbai, India, 2008, pp. 117– 120.

[5] Local and Metropolitan Networks — Part 16: Air Interface for Fixed Broadband Wireless Access Systems, IEEE Std. 802.16-2004, 2004.

For QPSK :

$$k_n = d \times i + j \quad for\,\forall\, j\ and\ for\,\forall\, i$$

For 16-QAM :

$$k_n = \begin{cases} d \times i + j & for\ j\%2 = 0\ or\ \forall\, i \\ d \times (1+1) + j & for\ j\%2 = 1\ and\ i\%2 = 0 \\ d \times (i-1) + j & for\ j\%2 = 1\ and\ i\%2 = 1 \end{cases}$$

For 64-QAM :

$$k_n = \begin{cases} d \times i + j & for\ j\%3 = 0\ and\ for\,\forall\, i \\ d \times (i-2) + j & for\ j\%3 = 1\ and\ i\%3 = 2 \\ d \times (i+1) + j & for\ j\%3 = 1\ and\ i\%3 \neq 2 \\ d \times (i+2) + j & for\ j\%3 = 2\ and\ i\%3 = 0 \\ d \times (i-1) + j & for\ j\%3 = 2\ and\ i\%3 \neq 0 \end{cases}$$

**Fig-2.** Algorithm to bypass the Floor function

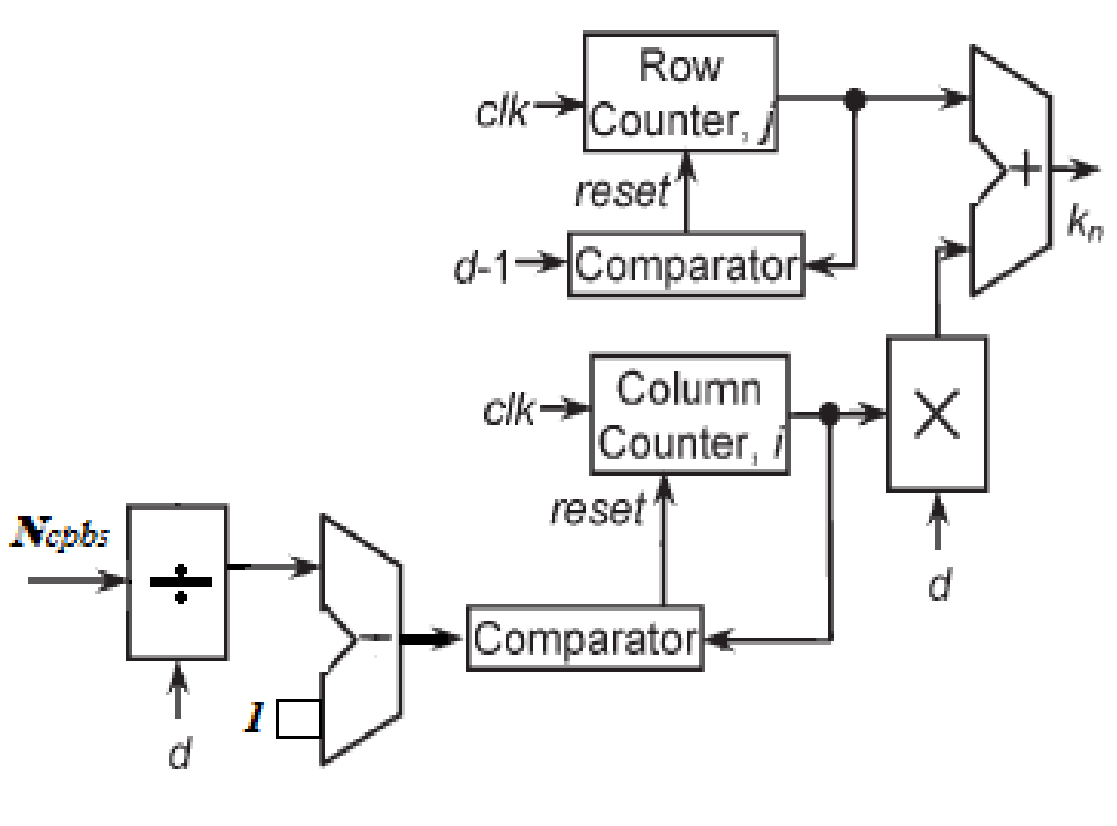

Fig. 3: QPSK Block Diagram

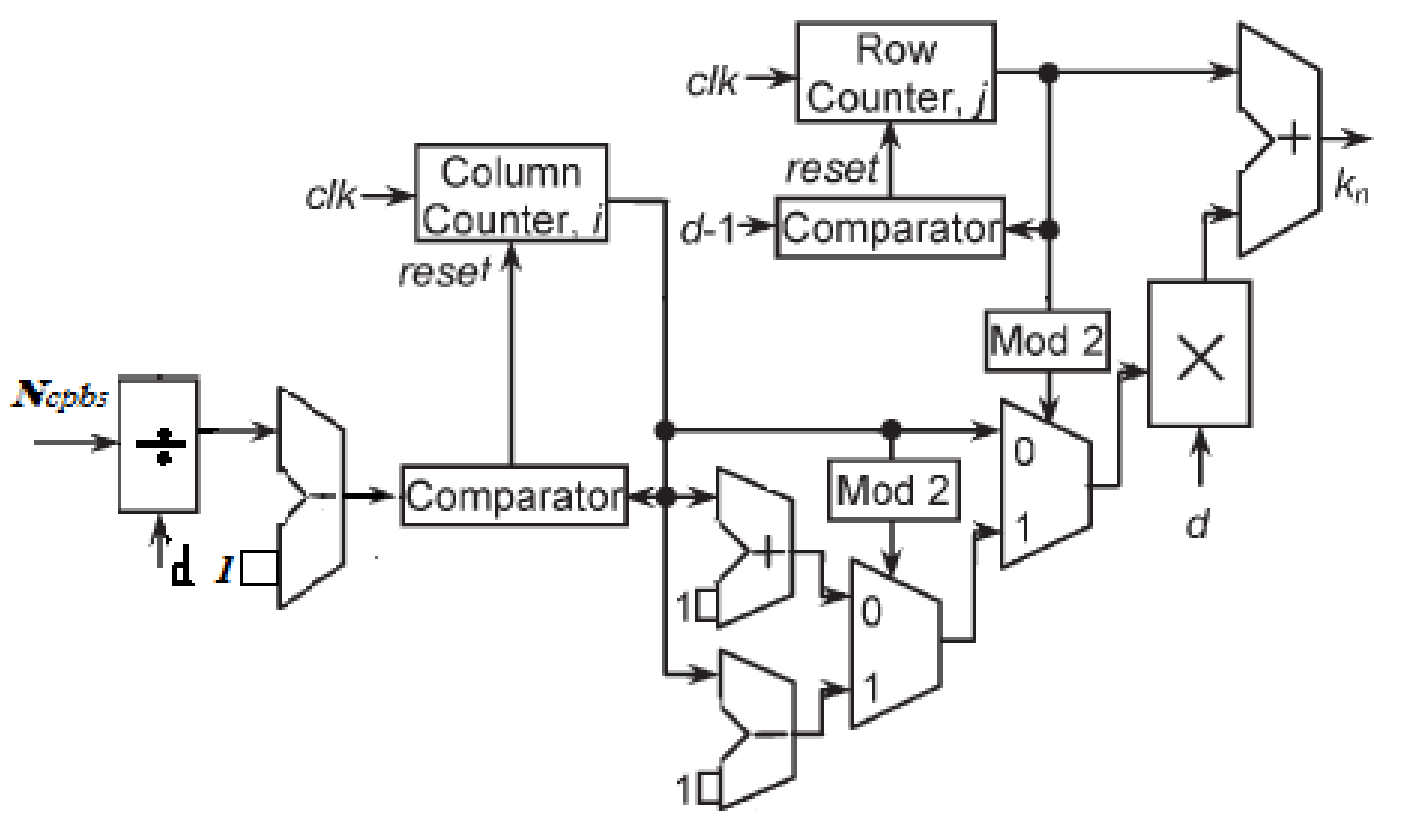

Fig. 4: 16-QAM Block Diagram

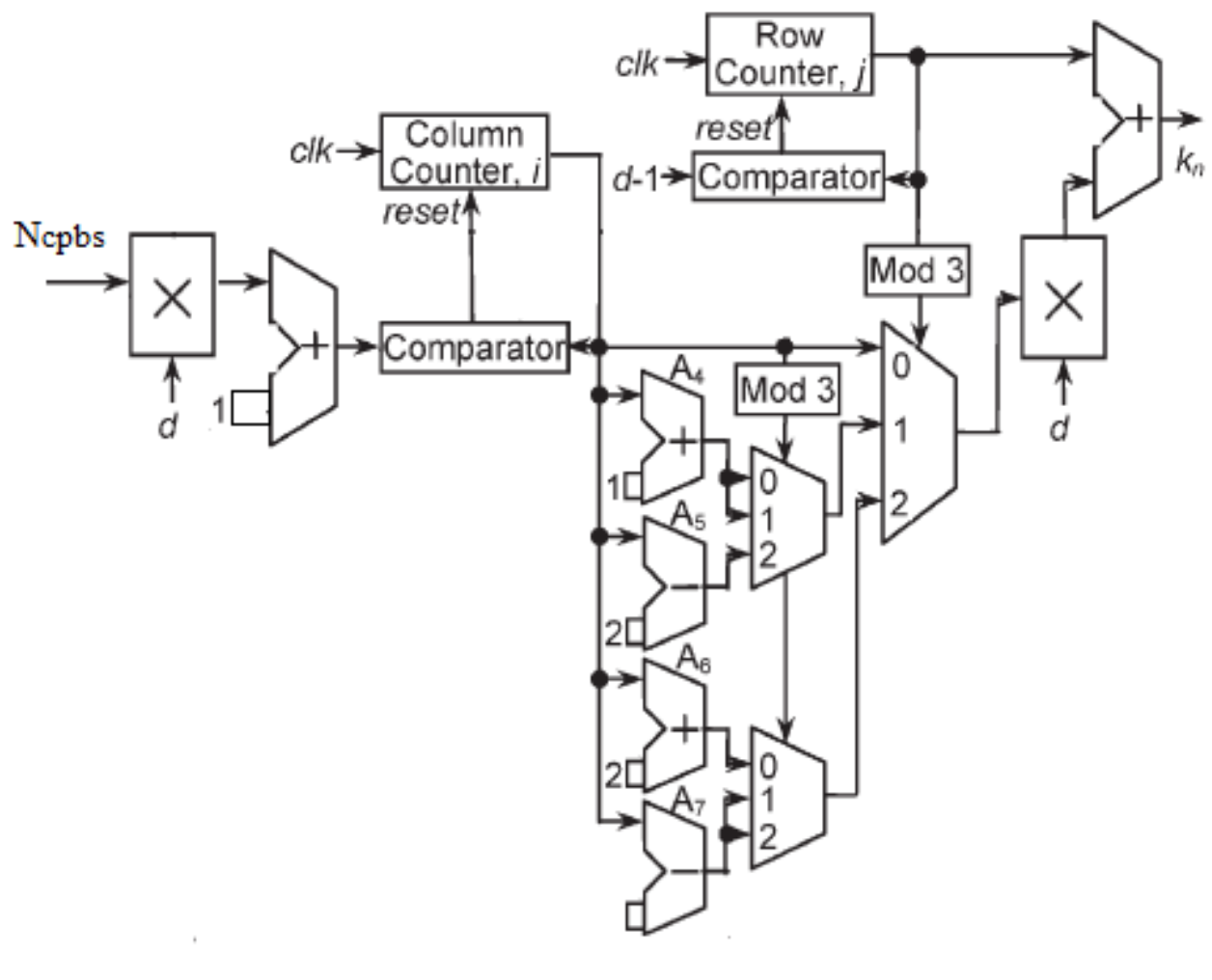

Fig. 5: 64-QAM Block Diagram

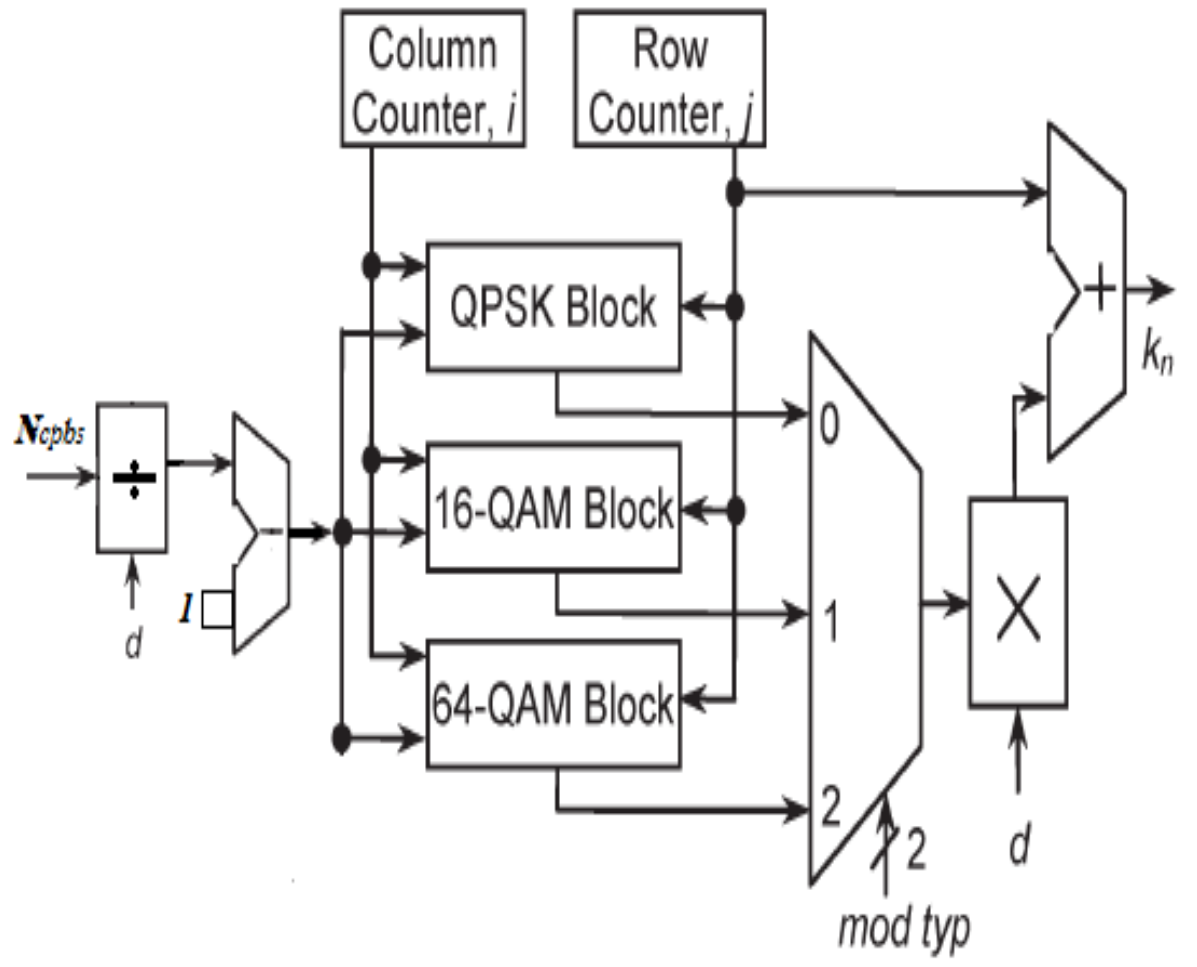

Fig. 6: Complete Deinterleaver Block Diagram

Table IV: Comparison between proposed and existing techniques.

| Device Specification | Results of Proposed Technique | Results of Upadhyaya & Sanyal | Performance of LUT based technique | % Reduction w.r.t. Upadhyaya & Sanyal | Remarks |
|---|---|---|---|---|---|
| **Slices** | 1% | 3.49 % | 17.66% | - 71.34 | Decrease |
| **Flip Flops** | 0.153% | 0.50 % | 0.78% | -69.4 | Decrease |
| **4 Input LUTs** | 1% | 3.35 % | 17.75% | -70.14 | Decrease |
| **Operating Frequency** | 130.24MHz | 121.82 MHz | 62.51 MHz | +6.9 | Increase |

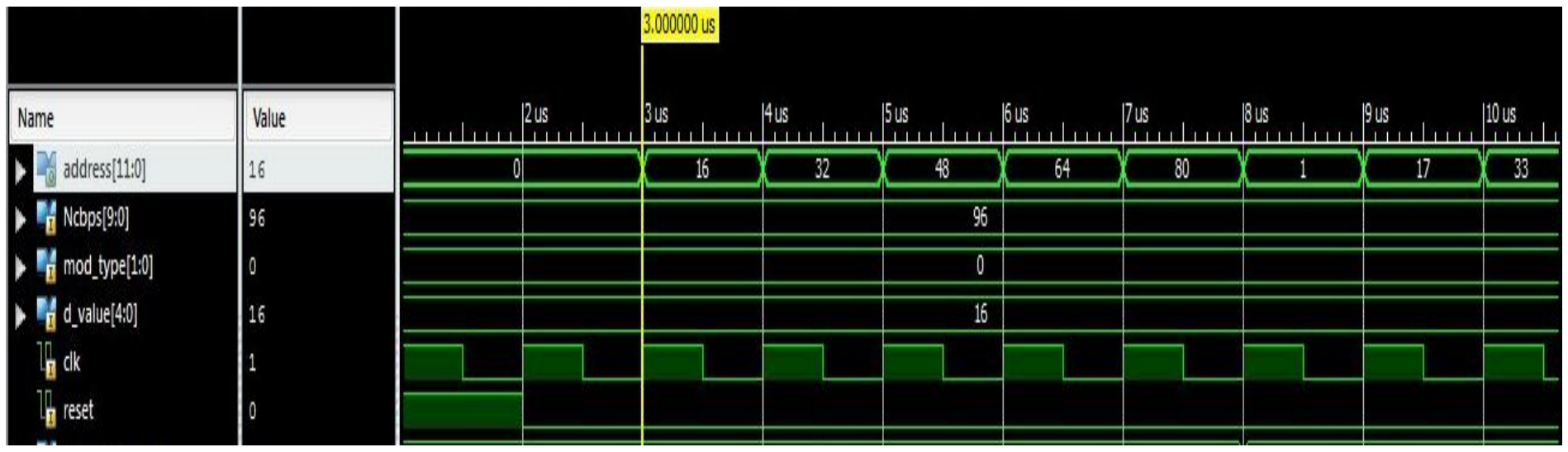

Fig 7: Initial addresses for Code Rate=0 and Ncpbs=96